\documentclass[letterpaper]{article} 
\usepackage{aaai23}  
\usepackage{times}  
\usepackage{helvet}  
\usepackage{courier}  
\usepackage[hyphens]{url}  
\usepackage{graphicx} 
\usepackage{natbib}  
\usepackage{multirow}
\usepackage{multicol}
\usepackage{booktabs} 
\usepackage{caption} 
\usepackage{algorithm}
\usepackage{algorithmic}
\usepackage{amsfonts,amssymb} 
\usepackage{amsmath}

\usepackage{color}

\usepackage{newfloat}
\usepackage{listings}
\DeclareCaptionStyle{ruled}{labelfont=normalfont,labelsep=colon,strut=off} 
\floatstyle{ruled}
\newfloat{listing}{tb}{lst}{}
\floatname{listing}{Listing}
\title{Toward Robust Diagnosis: A Contour Attention Preserving Adversarial Defense for COVID-19 Detection}
\author{
    Kun Xiang \textsuperscript{\rm 1}\equalcontrib,
    Xing Zhang \textsuperscript{\rm 2}\equalcontrib,
    Jinwen She,\textsuperscript{\rm 1}
    Jinpeng Liu,\textsuperscript{\rm 1}\\
    Haohan Wang,\textsuperscript{\rm 3}
    Shiqi Deng,\textsuperscript{\rm 1}
    Shancheng Jiang\textsuperscript{\rm 1}\textsuperscript{\rm 4}\thanks{Corresponding author.}
}
\affiliations{
    \textsuperscript{\rm 1}School of Intelligent Systems Engineering, Sun Yat-sen University\\
    \textsuperscript{\rm 2}Shuguang Hospital, Shanghai University of Traditional Chinese Medicine\\
    \textsuperscript{\rm 3}School of Information Sciences, University of Illinois Urbana-Champaign\\
    \textsuperscript{\rm 4}Guangdong Provincial Key Laboratory of Fire Science and Technology\\

    \{xiangk,shejw3,dengshq5\}@mail2.sysu.edu.cn, zhangxing2012@shutcm.edu.cn, liujp22@mails.tsinghua.edu.cn\\
    haohanw@illinois.edu, jiangshch3@mail.sysu.edu.cn
}

\usepackage{bibentry}

\begin{document}

\maketitle

\begin{abstract}
As the COVID-19 pandemic puts pressure on healthcare systems worldwide, the computed tomography image based AI diagnostic system has become a sustainable solution for early diagnosis. However, the model-wise vulnerability under adversarial perturbation hinders its deployment in practical situation. The existing adversarial training strategies are difficult to generalized into medical imaging field challenged by complex medical texture features. To overcome this challenge, we propose a Contour Attention Preserving (CAP) method based on lung cavity edge extraction. The contour prior features are injected to attention layer via a parameter regularization and we optimize the robust empirical risk with hybrid distance metric. We then introduce a new cross-nation CT scan dataset to evaluate the generalization capability of the adversarial robustness under distribution shift. Experimental results indicate that the proposed method achieves state-of-the-art performance in multiple adversarial defense and generalization tasks. The code and dataset are available at https://github.com/Quinn777/CAP.
\end{abstract}

\section{Introduction}

The coronavirus disease 2019 (COVID-19) pandemic has emerged as one of the pre-eminent global health crises of this century. After the earlier success in many fields\cite{ker2017deep, anwar2018medical,xiang2021novel}, various deep neural network-based frameworks (e.g. vision transformer) are proposed as primary screening tools for CT image-based medical diagnosis. However, recent studies found that these systems have exposed vulnerabilities under carefully crafted adversarial perturbations which is usually imperceptible to the human eyes \cite{szegedy2013intriguing}. It raises serious security concerns about their deployment, e.g., health insurance fraud and privacy disclosure, and even aggravating bias against AI-based Computer Aided Diagnosis system (CAD) \cite{mangaokar2020jekyll,finlayson2019adversarial}.

\begin{figure}[tbp] 
\centering 
\includegraphics[scale=0.7]{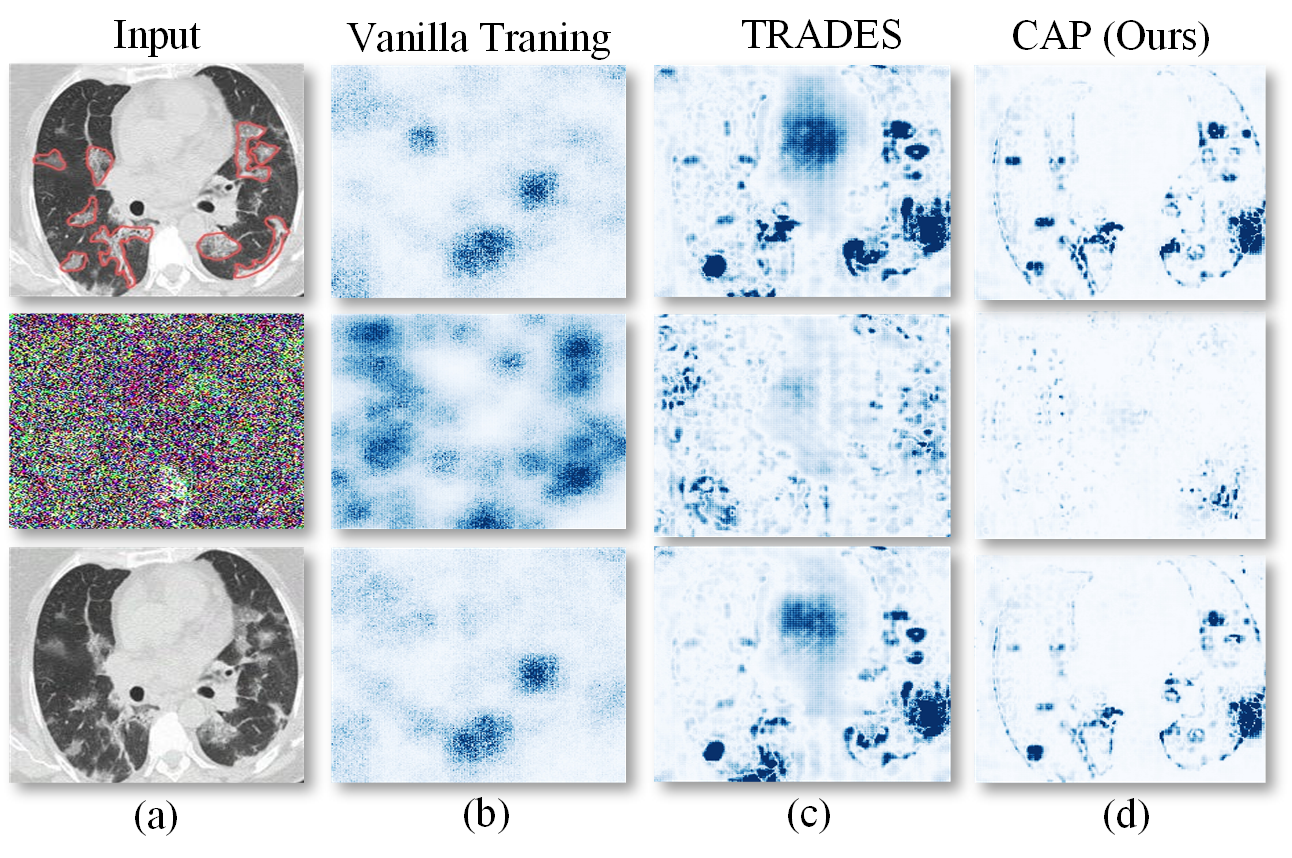} 
\caption{Saliency map \cite{simonyan2013deep} of three models under different training strategies. \textbf{Top}: Original COVID-19 positive CT scan with expert annotations. \textbf{Middle}: Adversarial noise generated by projected gradient descent (PGD). \textbf{Bottom}: Image under attack. All the models using \textbf{(b)} vanilla training, \textbf{(c)} TRADES and \textbf{(d)} CAP have correct prediction in clean samples, while only the last two models stay robust in adversarial case.} 
\label{img:visual} 
\end{figure}

To alleviate this problem, plenty of defense methods have been proposed, including well designed models \cite{zhou2022understanding}, detection techniques \cite{lee2018simple} and regularization strategies \cite{chan2019jacobian}, etc. Besides, adversarial training (AT) and its variants \cite{madry2017towards,wong2020fast,rade2021helper} stand out and become the powerful weapons to defend adversarial attacks. They generally produce artificial noise bounded in $\epsilon$-ball and formulate the key objective as a mini-max optimization problem. Despite their progress, due to the scarcity of medical images and the similarity of textures between classes, existing AT based methods are difficult to maintain attention to salient region or diagnostic points in medical imaging applications especially for diagnosis of COVID-19 based on CT scans. As shown in Figure \ref{img:visual}, model under vanilla training can only provide rough and incomplete information about the infection localization. Despite TRADES \cite{zhang2019theoretically}, as a representative AT method, improves the refined perception of local details, the model does not learn structured semantic features. For example, Figure \ref{img:visual}c focuses too much attention on the central spine which is unlikely to be infected.

In addition, it has been observed that there is a significant robustness generalization gap under data distribution shift \cite{rice2020overfitting,raghunathan2019adversarial}. Some studies provide a frequency-based explanation to reveal the unintuitive generalization behaviors between people and networks \cite{wang2020high}. Other findings try to improve the model input-robustness with out-of-distribution (OoD) generalization, e.g., data augmentation\cite{cubuk2018autoaugment}, feature mixup \cite{xu2020adversarial} and adversarial domain adaptation\cite{song2018improving}. However, existing adversarial robustness researches mainly focus on adapting the distribution discrepancy in natural image. In medical image data, the diverse distributions caused by different patients race, sampling equipment and post-processing methods, etc., usually result in a severer performance degeneration of models. Although the previous data driven methods introduce a larger data latent space, the excessive augmentation for lesion area could destroy the intrinsic structural information, such as lung outline and lesion margin. 

In this paper, we use the latest vision transformers as backbones and take the joint consideration of adversarial robustness and medical data distribution shift. A new medical benchmark data, including 874 COVID-19 and 873 NonCOVID-19 CT images across ten countries, is first established for validation. We then introduce a Contour Attention Preserving (CAP) framework for transferable adversarial training. In order to overcome the attention bias of models to infected areas, we intend to provide a spatial prior of lung cavity contour based on edge features extracted by self-guided filter \cite{he2010guided} transformation. We notice that the diagnostic features (such as diffuse patchy shadows and ground glass) are partially destroyed during filtering, which means the enhanced images should be separated from the original images and handled via different branches during the training process. Hence via surrogate model to indirectly minimize the prediction discrepancy between original and augmented counterparts, we generate a regularization term containing implicit prior information from the view of parameters. Without any labels, it is injected into the attention map of the training model in the form of parameter noises. Considering that a visual robust model should not rely on different metrics of distribution diversity, we then propose an optimization framework with hybrid distance metric. The framework uses different distance metrics in internal and external empirical risk calculations to correct the worst-case uncertainty. Figure \ref{img:visual}d shows the remarkable resolution of our method for scattered pulmonary nodules and provides lung contour information close to pixel level segmentation. In Section 4, several experimental evidences prove the effectiveness of our method. The main contributions of this paper are summarized as:
\begin{itemize}
    \item We propose an attention parameter regularization technique based on self-guided filtering. The regularization injects the prior knowledge of lung contour into vision transformers and introduce a soft consistency between natural images and augmentations.
    \item A transferable adversarial training framework, CAP, is indicated to defense the imperceptible perturbations in COVID-19 CT data. We raise a hybrid distance metric to further optimize the min-max problem.
    \item We propose a new CT scan benchmark, COVID-19-C, to evaluate the generalization capacity. Extensive experiments demonstrate that we consistently improve the adversarial robustness and interpretability of state-of-the-art method even under complex data distribution shift.
\end{itemize}

\begin{figure*}[th] 
\centering 
\includegraphics[scale=0.9]{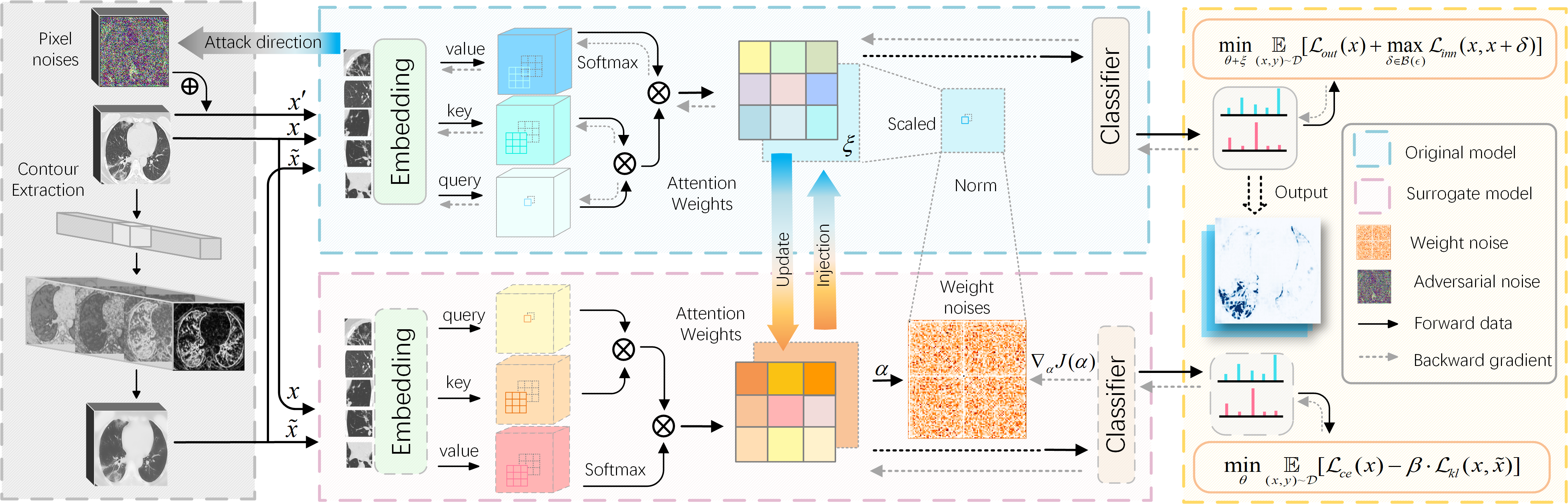} 
\caption{The overall framework of the proposed CAP. In the gray area of the left figure, we first employ a contour extraction module to obtain the prior information of lung cavity. Then, a surrogate model (the middle red area) generates attention noise to attack the weight of target model (the middle blue area). Finally, we optimize the adversarial loss with hybrid distance metric (the right yellow area).} 
\label{img:framework} 
\end{figure*}

\section{Related Work}
\subsection{Robustness under Distribution Shift}
A large number of works demonstrate the fragility of neural network in artificially corruptions and unseen data distributions \cite{geirhos2017comparing,hendrycks2018benchmarking}. \cite{wang2019learning} theoretically show global representation can activate more domain-specific signals in the process of gradient propagation. \cite{huang2020self} applies a feature-critic training heuristic to enforce model's tendency in predicting with none-predictive features. Another group of methodologies to tackle generalization problem are based on data augmentation. \cite{zhang2017mixup} creates mixture samples via different inputs with soft labels, while CutMix \cite{yun2019cutmix} randomly cuts out samples before mixing. Several works provide a new Fourier-based perspective to fit the data distribution bias \cite{chen2021amplitude,xu2021fourier}. Furthermore, regularization loss is also widely used for keeping the consistency of predictions between original and augmented inputs\cite{wang2022toward,abuduweili2021adaptive}. But for texture-sensitive medical imaging task, these strong regularization constraints will inject inductive bias especially with a excessive augmentation method.

\subsection{Robustness under Adversarial Attacks}
Robust training method has experienced explosive development to address the problem of adversarial deception \cite{szegedy2013intriguing, goodfellow2014explaining}. Powerful defense ability is obtained by minimizing the training loss, e.g., gradient based methods  \cite{athalye2018obfuscated, kurakin2018adversarial}, carefully designed training strategies \cite{xie2020self, deng2021improving} and self-attention based networks \cite{benz2021adversarial, paul2022vision}. In medical vision tasks, it is proved that AI systems show less input-robustness than what they have achieved in natural image data \cite{ma2021understanding,kaviani2022adversarial}. \cite{hirano2021universal} find CNNs are vulnerable to nontargeted universal perturbation and provide simple detectors to search the feature difference. \cite{lal2021adversarial} develops a feature fusion technique to defense speckle noise attacks in riabetic retinopathy recognition task while other work improve robustness for 3D whole heart segmentation \cite{liao2020mmtlnet}. However, there are few works study the adversarial robustness on COVID-19 CT scans, which present less obvious characteristic patterns. Besides, the adversarial robustness toward distribution shift is not taken into account in previous studies, which causes damage to universality and safety of AT methods in practical deployment.

\section{Methodology}
\subsection{Definition and Overview}
Formally, for a given data point $(x_i,y_i)$ sampled from current $C$-class image dataset $D\subseteq(\mathcal X,\mathcal Y)$, its augmentation is denoted as $(\tilde x_i,\tilde y_i)\sim \tilde D$ where $i\in \lbrace1,...,N\rbrace$ is the number of training examples. We represent $f$ to be the discriminative model parameterized by $\theta$, which can be perturbed by attention weight noise $\xi$. The input adversarial noise is denoted as $\delta$ which is bounded by $||\delta||_p \le \epsilon$.

In this section, we first employ a self-guided filter based transformation for edge  extraction. The prior information of lung cavity is injected via a well designed attention parameter regularization method. Then we integrate hybrid distance metrics to formulate the empirical risk of min-max optimization. The diagram of our method is shown in Figure 2. We now discuss the technical details of our method.

\subsection{Contour-preserving Data Augmentation}
We first consider to leveraging strong prior knowledge acquired from practical diagnostic experience. Since the tissue lesion caused by COVID-19 often occurs in the interior of lung, we intend to extract this position correlation from lung cavity contour feature. Inspired by the previous work, Guided Filter \cite{he2010guided}, we adopt a local linear augmentation to smooth small fluctuations without losing edge details. We assume the guidance image in pixel $j$ is $I^{(j)}$ and the filtered output image is $\tilde{x}$. The transformation can be formulated as:
\begin{equation}
\begin{split}
\label{eq:self-guided}
\tilde{x}^{(j)}=\bar{a}_k x^{(j)}+\bar{b}_k, \ \ \forall{j}\in \omega_k
\end{split}
\end{equation}
where the $k$ is the position index of a local rectangle filter window $\omega$ with size $s\times s$. $\bar{a}_k$ and $\bar{b}_k$ are the average of linear coefficients $a_k$ and $b_k$ respectively, which can be optimized by minimizing the difference between $\tilde{x}^{(j)}$ and the input pixel $x$. In order to avoid introducing deviations outside the current class, we assume the natural input image $x$ as the self-guidance term $I$. The solution of the above linear ridge regression is given as:
\begin{equation}
\begin{split}
\label{eq:self-guided2}
&a_k=\frac{\frac{1}{|\omega|}\sum_{i\in \omega_k}x_i^2-\mu_k \bar x_k}{\sigma^2_k+t}, \ b_k=\bar{x}_k-a_k\mu_k
\end{split}
\end{equation}
Here $\mu_k$ and $\sigma_k$ are mean and variance of $I$ in the window $k_{th}$. $\bar x^{(k)}$ is the average of $x$ in $\omega_k$ and we use a scaling temperature $t$ to control the degree of smoothness. Since the filter is a edge-preserving function, the machine noise and ground glass in original scans will be dropped and there only left a few trachea tissues and pulmonary nodules. Compared with the frequency domain based and interpolation based methods\cite{chen2021amplitude,zhang2017mixup}, we highlight the semantic structures and global visual cues.

\subsection{Attention Parameter Regularization}

The contour-preserving data augmentation allows us to get prior information about the approximate location of lesions. However, simply adopting consistency regularization may neglect the condition that intrinsic information of the instance could be moved during augmentation. As a result, inspired by previous works\cite{wu2020adversarial,he2019parametric,wen2018flipout}, we propose an attention parameter regularization mechanism with local equivariance to exert a softer constraint. The regularization generates negative weight noise from enhanced images and implicitly introduces the prior knowledge of organ location to attention map.

Considering a simple vision transformer, we add the noise only to the multi-headed self-attention (MSA) module which focuses on the positional relationship between patches. Considering the edge preserving ability of guided filtering, the distributions between the original and augmented images have a extensive similarity. We expect the decision boundary toward both two types of samples to be close to maintain a implicit consistency. At the same time, the noise confuse the prediction of natural samples by enlarging the expectation of Wasserstein distance between conditional distributions for each clean class. We utilise a mirror image of current model as a surrogate and the surrogate loss is constructed for maximizing this diversity:
\begin{equation}
\begin{split}
\label{eq:surrogate loss}
\mathcal{L}_{sur}=\mathop{\mathbb{E}}_{(x,y)\sim\mathcal{D}}[\mathcal{L}_{ce}(x,y;\theta)-\beta \cdot\mathcal{L}_{kl}((x,\tilde x), y;\theta]
\end{split}
\end{equation}
where $\mathcal{L}_{kl}$ is Kullback–Leibler divergence and $\mathcal{L}_{ce}$ is cross-entropy loss function. We re-write the weight of attention layer as $\alpha$. Then with a batch size $m$, the gradient of surrogate attention weight can be calculated by: 
\begin{equation}
\begin{split}
\label{eq:gradient}
\nabla _{\alpha}J(\alpha)=\frac{1}{m}\sum_{i=1}^{m}\nabla_{\alpha} \mathcal L_{sur}((x_i,\tilde x_i),y_i;\alpha)
\end{split}
\end{equation}
To overcome the numerical instability problem, we adopt gradient normalization and control the perturbation amplitude with scaled ratio $\gamma$. The final parameter noise can be formulated by:
\begin{equation}
\begin{split}
\label{eq:attention noise}
\xi=\eta \frac{\nabla_\alpha J(\alpha)}{||\nabla_\alpha J(\alpha)||_2}\cdot\gamma ||\alpha||_2
\end{split}
\end{equation}
Eq. \ref{eq:surrogate loss}, \ref{eq:gradient}, \ref{eq:attention noise} are all computed in the surrogate model. Similar with \cite{wu2020adversarial}, we move the attention weight back to the center with the same step size after back propagation. Then the final gradient update after simplification can be expressed as:
\begin{equation}
\begin{split}
\label{eq:gradient update}
\theta \gets &\theta - \frac{\eta}{N}\sum_{i=1}^{N}\nabla_{\theta+\xi} \mathcal L_{out}((x_i,x_i+\delta),\hat y_i;\theta+\xi) \\
&where \ \ \delta=\mathop{\arg\max}\limits_{\delta\in\mathcal{B}(\epsilon)}\mathcal{L}_{inn}(x_i+\delta,y_i;\theta+\xi)
\end{split}
\end{equation}
 Note that in Eq. \ref{eq:gradient update}, the outer label $\hat y_i$ in $\mathcal L_{out}$ is softed by $\hat y = \epsilon \cdot f(\tilde x_i) +(1-\epsilon) \cdot y_i$. Since $\xi$ is directly proportional to $\mathcal{L}_{sur}$, weight perturbation is added to original $\theta$ in the process of adversarial learning for maximizing the surrogate error. The weight perturbation implicitly introduces a consistency constraint and narrows the distributions between original and augmented images. Besides, since the weight noise is negative and it forces the model to make wrong decisions for both two types of inputs. The intuition behind is that, the gradient-descent optimizer attempt to find a equilibrium point that perceive model-wise worst-case and keep contour consistency.

\subsection{Hybrid Distance Metric based Optimization}
In CAP, we improve the adversarial loss function with hybrid distance metric to substitute $\mathcal{L}_{inn}$ and $\mathcal{L}_{out}$ in Eq. \ref{eq:gradient update}. Previous studies have shown that a metric satisfying the distance axioms can help regulate the input inductive bias when maximizing the between class distances\cite{pang2022robustness,wang2020once}. Considering a well-trained model that achieves high accuracy on both natural and adversarial data distributions, we believe that the input robustness should not completely depend on how the model measure a specific distribution shift in the adversarial training process. Starting from this point, we attempt to make some changes to original adversarial framework with different distance metrics in CAP framework, with purpose of relieving the inductive bias in optimization. 

We denote the class posterior probability as $p(y_i|x_i)$. Following \cite{golik2013cross}, the square error can be written by:
\begin{equation}
\begin{split}
\label{eq:se}
\mathcal L_{se}=\sum_i^N\sum_c^C[(p(y_k|x_i)-\psi(y_k,y_i)]^2
\end{split}
\end{equation}
where $\psi(\cdot,\cdot)$ is Kronecker function. Compared with KL divergence, square error does not introduce the negative logarithm of posterior probability. Thus there is a smooth transition in the training process. However, our empirical observations demonstrate that when the output is close to the optimal or the worst, the activation value tends to be saturated which manifests a flat loss plane, even though learning is not over. In particular, since the step size and magnitude of perturbation are usually small while generating strong rival classes in the update of adversarial attacks, it brings more difficulties to converge to the farthest position when facing gradient vanishing. Consequently, we consider applying square error function only in external minimization problem as $\mathcal L_{out}$. Our final optimization objectives can be summarized as follows:

\begin{equation}
\begin{split}
\label{eq:objective}
\mathop{\min}\limits_{\theta+\xi}\mathop{\mathbb{E}}_{(x,y)\sim\mathcal{D}}
&[\mathcal{L}_{se}(x,y;\theta+\xi)+\beta \cdot \mathcal{L}_{se}(x,x+\delta;\theta+\xi)]\\
&where \; \delta=\mathop{\arg\max}\limits_{\delta\in\mathcal{B}(\epsilon)}\mathcal{L}_{kl}((x,x+\delta),y;\theta+\xi)
\end{split}
\end{equation}

We also provide some theoretical evidences to bound the formulation. we remove $\mathcal{L}_{se}(x_i,y_i)$ in sample $x_i$ with label $y_i$ by:

\begin{equation}
\begin{split}
\label{eq:se2}
\sum_i^N[p(y|x_i)-\psi(y,y_i)]^2=[1-p(y_i|x_i)]^2+\sum_{y_i\ne y}q^2(y|x_i)
\end{split}
\end{equation}

As such, we substitute it into Eq. \ref{eq:se2} to bound the external optimization:
\begin{equation}
\begin{split}
\label{eq:bound}
&0 \le \mathcal{L}_{out}(x_i,y_i) \le 2+2\beta
\end{split}
\end{equation}

Here $\mathcal{L}_{out}$ is the external loss in Eq. \ref{eq:objective}. Detailed proof is shown in Appendix A. Note that the notation $x_i'$ is used as an abbreviation for adversarial sample $x_i+\delta_i$. The square loss reaches the upper bound when there is a strong rival class and the decision boundary confuse both natural and adversarial samples. We provide some empirical evidence for the above theory in Table \ref{tbl:distance}. The compound metrics significantly improve the performance in both TRADES and CAP frameworks. Our final pipeline is summarized in Figure \ref{img:framework}.

\begin{figure*}[htb] 
\centering 
\includegraphics[scale=0.45]{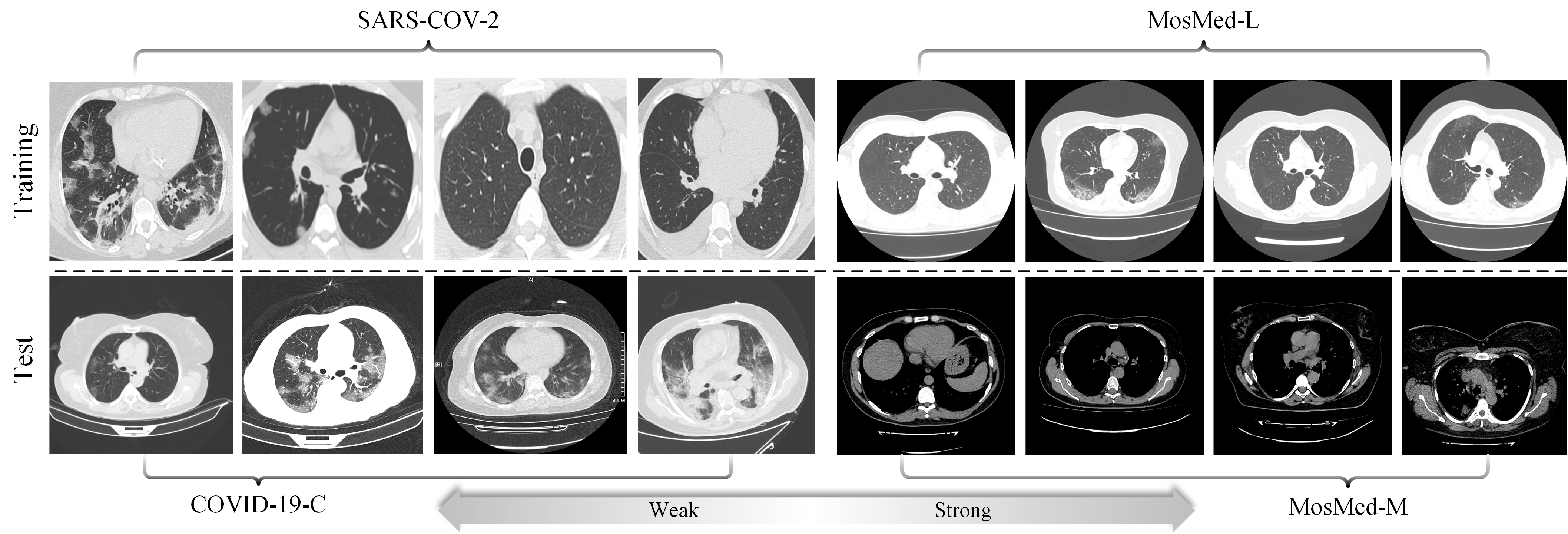} 
\caption{Four datasets we used for evaluation.The test sets include images with both weak (COVID19-C) and strong (MosMed-M) distribution shift.} 
\label{img:datasets} 
\end{figure*}

\begin{table}[H]
\centering
\caption{Effect of different distance metrics on optimization problems. Outer/Inner represent the distance metric used in minimum/maximum optimization problem, respectively.}
\scalebox{0.9}{
\begin{tabular}{c|cc|cc}
\toprule
        Method & Outer & Inner & Clean & PGD-10\\
        \midrule
        \multirow{4}*{TRADES} & KL & KL & 88.73 & 77.67\\
        & SE & SE & 84.51 & 41.85\\
        & KL & SE & 84.50 & 50.70\\
        & SE & KL & \textbf{91.95} & \textbf{84.31}\\
        \midrule
        \multirow{4}*{CAP (Ours)} & KL & KL & 86.72 & 58.35\\
        & SE & SE & \textbf{93.96} & 63.68\\
        & KL & SE & \textbf{93.96} & 41.85\\
        & SE & KL & \textbf{92.96} & \textbf{85.92}\\

\bottomrule
\end{tabular}
}
\label{tbl:distance}
\end{table}

\section{Experiments}

\paragraph{Datasets}
We use two public COVID-19 CT databases for evaluation, i.e., \textbf{SARS-COV-2} \cite{morozov2020mosmeddata} and \textbf{MosMed} \cite{soares2020sars}. SARS-COV-2 contains 2482 2D lung window images, including positive and negative patients. MosMed provides a total of 1110 3D scans, which is used for infection severity quantification. It contains 5 types of labels (healthy, mild, moderate, severe and critical), which are subdivided by the infection percentage of CT volume. We adopt Hounsfield transformation to convert it into two modalities of CT images, i.e., lung window and mediastinal window. They are named as \textbf{MosMed-L} and \textbf{MosMed-M} respectively.

To evaluate the generalization capacity under distribution shift, we establish a comprehensive database named \textbf{COVID19-C}. COVID19-C is collected from mutiple public datasets \cite{wang2020covid,hemdan2020covidx,morozov2020mosmeddata,soares2020sars} and involves 1747 images from patients in 10 different countries. Each case is chosen by experienced radiologists to cover as much varieties from different background and texture shift as possible. We divide COVID19-C into 4 subsets according to regions, which are abbreviated as\textbf{ COV-C} (China), \textbf{COV-R} (Russia), \textbf{COV-I} (Iran) and \textbf{COV-M}. Note that COV-M contains patients from multiple countries and samples in four subsets are independent. Examples are shown in Figure. \ref{img:datasets} and Table. \ref{tbl:dataset}.

\paragraph{Settings}
The proposed method is compared with a series of AT and OoD baselines, i.e., FAST \cite{wong2020fast}, AT \cite{madry2017towards}, TRADES \cite{zhang2019theoretically}, AVmixup \cite{lee2020adversarial}, HAT \cite{rade2021helper}, FAT \cite{zhang2020attacks}, AWP \cite{wu2020adversarial}, MMA \cite{ding2018mma}, SCORE \cite{pang2022robustness}; Mixup \cite{zhang2017mixup}, AutoAug \cite{cubuk2018autoaugment}, CutOut \cite{devries2017improved}, CutMix \cite{yun2019cutmix}, FACT \cite{xu2021fourier} and APRS \cite{chen2021amplitude}. For SARS-COV-2, we use Visformer-Tiny and Deit-Tiny as our backbones and train them for 200 epochs. We choose AdamW optimizer with cosine annealing learning rate 0.0005 for Visformer while 0.0001 for Deit. For MosMed-L, we adopt Visformer-Tiny with AdwmW, learning rate of 0.0002 with 20 steps decay. Models are trained for 200 epochs with batch size 64. Attacks are conducted for 10 steps in SARS-COV-2 (5 steps in MosMed-L) with step size of $2/255$ and $l_{\infty}$ perturbation budget $\epsilon=8/255$. In CAP, we set scaled ratio $\gamma=0.0001$, $t=0.003$ with kernel size 5. Training and test sets are divided by 8:2 and images are resized to $224\cdot224$. We implement experiments with PyTorch framework on Nvidia RTX 3090 GPUs and more details are shown in Appendix C.

\begin{table}[thb]
\centering
\caption{Datasets for COVID-19 binary detection task.}
\scalebox{0.68}{
\begin{tabular}{c|ccccccc|c}
\toprule
Dataset & Class & Positive & Negative & Total & Country & Slice Selection\\
\midrule
SARS-COV-2 & 2 & 1252 & 1229 & 2481 & Brazil & Expert \\
\midrule
COV-C & 2 & 1429 & 1648 & 3077 & China & Expert \\
COV-R & 2 & 236 & 702 & 938 & Russia & Expert \\
COV-I & 2 & 105 & 105 & 210 & Iran & Expert \\
COV-M & 2 & 182 & 170 & 352 & - & Automatic\\
\bottomrule
\end{tabular} 
}
\label{tbl:dataset}
\end{table}

\begin{figure}[htb] 
\centering 
\includegraphics[scale=0.7]{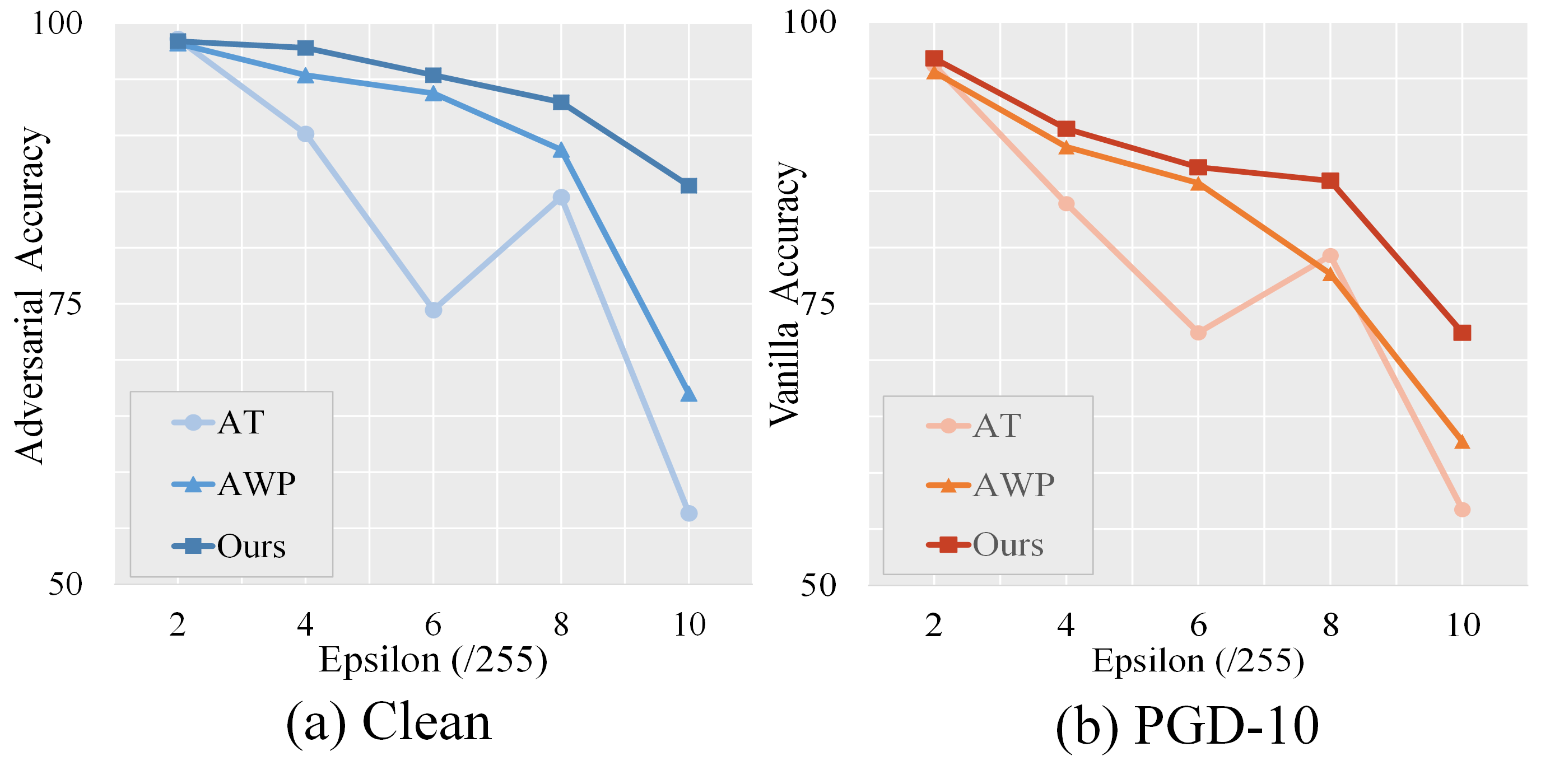} 
\caption{Effect of a wide range of perturbation budget ($\epsilon$) in adversarial training procedure with different method on SARS-COV-2.} 
\label{img:different perturbation} 
\end{figure}

\begin{table*}[h]
\centering
\caption{Clean accuracy, adversarial accuracy and OoD adversarial accuracy ($\%$) on SARS-COV-2 and COVID-19-C. We perturb OoD data with PGD-10 and all the attacks are generated with 10 iterations ($\epsilon=8/255$ and step size $2/255$).}
\scalebox{0.8}{
\begin{tabular}{cc|cccccc|ccccc}
\toprule
\multicolumn{13}{c}{Visformer-Tiny}\\
\midrule
Method & Clean & FGSM & R-FGSM & PGD &MIM & AutoAttack & Avg & COV-C & COV-R & COV-I & COV-M & Avg\\
\midrule
FAST&81.69&81.49&79.88&76.26&72.84&72.43&76.58&32.98&27.84&46.67&43.51&37.75\\
AT&84.51&79.68&82.29&79.28&79.48&79.28&80.00&42.23&44.03&46.19&48.97&45.36\\
TRADES&88.73&80.08&84.31&77.67&78.74&74.25&79.01&44.24&45.74&49.05&47.38&46.60\\
AVmixup&\textbf{93.16}&77.67&85.71&74.85&75.25&74.25&77.55&30.97&38.07&31.43&42.37&35.71\\
HAT&84.71&67.81&76.06&67.27&67.61&66.6&69.07&40.35&39.77&23.33&38.50&35.49\\
FAT&92.76&82.09&87.73&81.09&81.29&80.89&82.62&39.01&48.58&38.10&47.15&43.21\\
AWP&90.54&80.08&84.51&77.87&78.47&76.26&79.44&41.55&44.32&\textbf{50.00}&49.89&46.44\\
MMA&75.25&63.38&67.81&61.17&60.97&57.34&62.13&30.56&40.91&46.67&46.47&41.15\\
SCORE&84.51&42.05&42.05&41.85&41.65&15.49&36.62&37.67&36.65&30.48&44.65&37.36\\
TRADES+Mixup&82.49&58.55&70.62&56.34&56.54&56.34&59.68&30.97&38.07&15.24&38.72&30.75\\
TRADES+AutoAug&71.63&65.39&67.81&64.79&64.99&64.39&65.47&32.17&39.49&42.38&43.05&39.27\\
TRADES+CutOut&85.51&75.86&78.04&73.44&68.01&62.78&71.63&31.37&37.78&46.46&46.47&40.52\\
TRADES+CutMix&76.66&56.34&64.99&53.92&54.33&53.52&56.62&20.51&24.15&31.9&21.64&24.55\\
TRADES+FACT&84.10&78.07&80.28&77.46&77.46&77.46&78.15&39.54&44.32&35.24&39.41&39.63\\
TRADES+APRS&87.32&76.66&82.49&75.05&75.65&74.45&76.86&38.20&32.10&43.33&45.79&39.86\\
\midrule
CAP(Ours)&92.96&\textbf{86.52}&\textbf{90.54}&\textbf{85.92}&\textbf{85.92}&\textbf{85.92}&\textbf{86.96}&\textbf{45.58}&\textbf{50.57}&48.10&\textbf{53.99}&\textbf{49.56}\\
\midrule
\midrule
\multicolumn{13}{c}{Deit-Tiny}\\
\midrule
Method & Clean & FGSM & R-FGSM & PGD &MIM & AutoAttack & Avg & COV-C & COV-R & COV-I & COV-M & Avg\\
\midrule
FAST&93.76&60.36&75.25&39.44&47.48&23.54&49.21&8.85&5.68&0.01&2.73&4.32\\
AT&91.55&83.5&87.12&81.49&82.09&81.09&83.06&35.79&\textbf{44.03}&46.19&39.18&41.30\\
TRADES&94.37&84.91&89.34&82.49&83.10&82.09&84.39&36.73&40.91&44.76&46.01&42.10\\
AVmixup&94.37&85.51&90.54&82.70&82.90&82.29&84.79&36.86&40.91&44.29&45.10&41.79\\
HAT&95.17&82.70&89.94&79.68&80.28&79.07&82.33&26.14&21.59&38.57&20.73&26.76\\
AWP&94.16&85.11&89.94&82.70&83.10&81.09&84.39&39.01&43.18&44.76&45.10&43.01\\
MMA&83.50&68.01&73.84&64.19&64.99&64.19&67.04&30.29&23.86&44.76&35.99&33.73\\
SCORE&82.09&73.74&77.06&60.36&55.73&36.42&60.66&35.25&38.07&12.86&32.35&29.63\\
\midrule
CAP (Ours)&\textbf{95.57}&\textbf{87.12}&\textbf{90.74}&\textbf{84.31}&\textbf{85.51}&\textbf{83.70}&\textbf{86.28}&\textbf{44.88}&42.90&\textbf{47.62}&\textbf{47.38}&\textbf{45.70}\\
\bottomrule
\end{tabular}
}
\label{tbl:sars}
\end{table*}

\section{Results}
\subsection{COVID-19 Detection on SARS-COV-2}
\paragraph{White-box Attack}
To evaluate the performance of our proposed method, we first employ a wide range of white-box adversarial attacks with $l_\infty$ threat model: FGSM \cite{goodfellow2014explaining}, R-FGSM \cite{tramer2017ensemble}, PGD \cite{madry2017towards}, MIM \cite{dong2018boosting} and Auto Attack\cite{croce2020reliable}. As reported in Table \ref{tbl:sars}, we achieve the highest adversarial accuracy in Visformer under all five attacks and improve the SOTA result (FAT) by 4.34$\%$ on average while maintaining natural accuracy. A narrower gap between clean and adversarial accuracy is observed in CAP (6$\%$ vs 10.14$\%$ compared with FAT). In Deit, we get 1.2$\%$ and 1.49$\%$ improvement than AVmixup in clean and adversarial samples, respectively. Moreover, Figure \ref{img:different perturbation} shows CAP consistently improves the robustness with different perturbation budgets.

\begin{figure*}[th] 
\centering 
\includegraphics[scale=0.5]{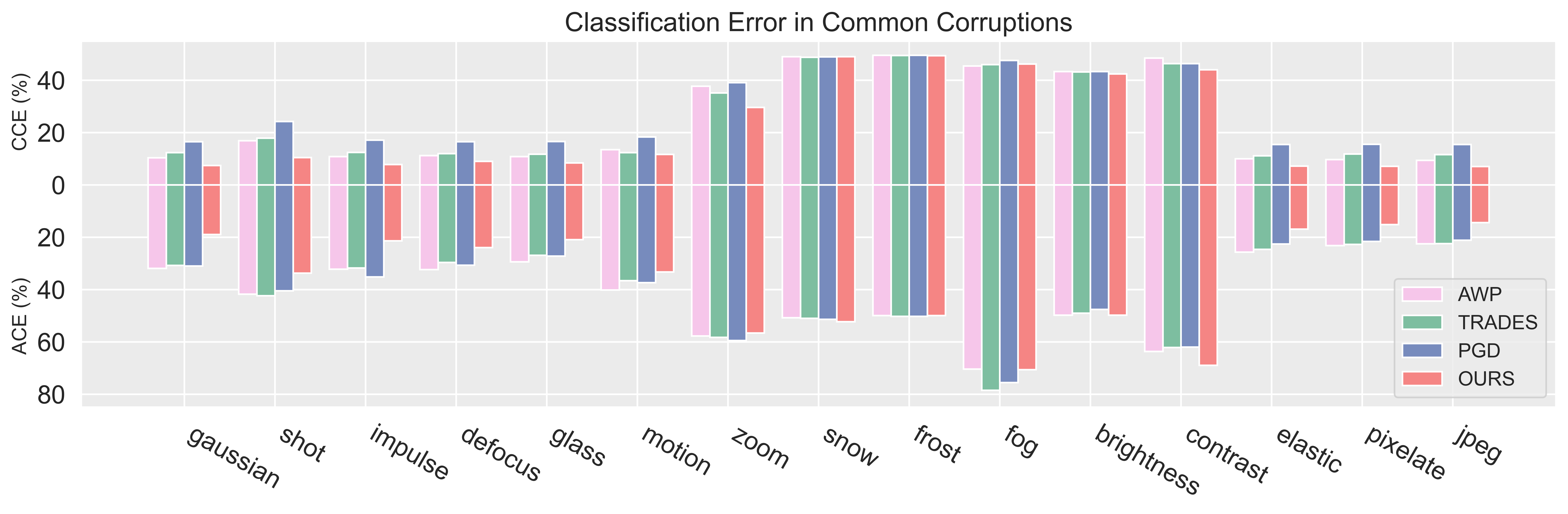} 
\caption{Clean Corruption Error (CCE) and Adversarial Corruption Error (ACE) for different methods with Visformer-Tiny in corrupt SARS-COV-2.} 
\label{img:mce} 
\end{figure*}

\begin{figure}[htb] 
\centering 
\includegraphics[scale=0.48]{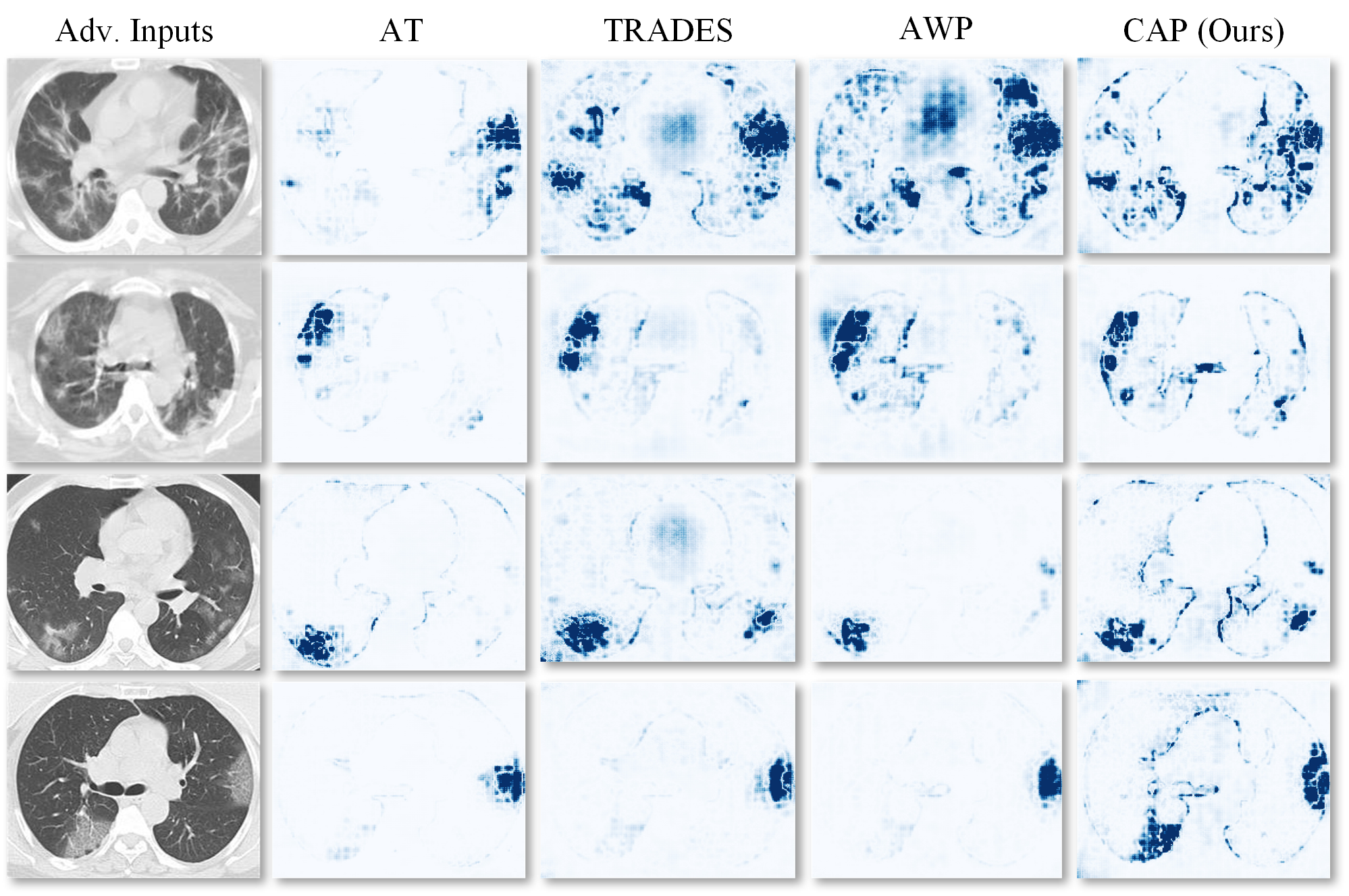} 
\caption{Saliency maps on models with different AT methods. The input images are perturbed by PGD-10 with $\epsilon=8/255$.} 
\label{img:interpretability} 
\end{figure}

\paragraph{Distribution Shift}
In addition to surpassing numerous previous methods in source data distribution, from Table \ref{tbl:sars}, it is evident that CAP also outperforms them under distribution shift. We introduce COVID19-C for generalization ability evaluation and images are perturbed by PGD-10 with $\epsilon=8/255$ and step size $2/255$. Results demonstrate that we gain an average improvement of 3.12$\%$ and 2.69$\%$ than SOTA baselines with Visformer and Deit respectively. More intriguingly, existing augmentations, e.g., Cutout (9.04$\%$ lower than ours) and AutoAug (10.29$\%$ lower than ours), cannot transfer adversarial robustness to unseen distribution as they do in natural images. Furthermore, the means±S.D of clean accuracy, average adversarial accuracy and OoD adversarial accuracy for Visformer are listed as follows ($\%$): $\{$82.13±9.89, 70.10±11.91, 38.91±5.68$\}$ and $\{$91.12±4.92, 74.48±12.86, 32.83±12.26$\}$ for Deit. The results show that the SOTAs in Deit achieve better adversarial robustness while bring larger generalization error. Overall, the experimental results prove that our enhancement avoids excessive destruction of structural information and accurately represents the ontology semantics.

\paragraph{Common Corruption}
Due to the discrepancy in hardware and shooting parameters, there exists surface variations between input scans. Hence we evaluate the generalization behavior with 15 common corruptions \cite{hendrycks2019benchmarking}, e.g., gaussian noise and elastic. All the models are tested with CCE and ACE, which can be calculated by $CCE_c=\frac{1}{5}\sum_{s=1}^5 E_{s,c}(x)$ and $ACE_c=\frac{1}{5}\sum_{s=1}^5 E_{s,c}(x')$. Here $x'$ is corrupt image with PGD attack like Table \ref{tbl:sars}. Figure \ref{img:mce} shows our CAP decreases both CCE and ACE from SOTA method (AWP) in most corruptions, e.g. shot noise CCE $10.42\%$ vs $16.86\%$ and ACE $33.76\%$ vs $41.77\%$. The mean CCE and mean ACE in CAP is $22.42\%$ and $36.48\%$, which are both better than AWP ($25.06\%$ and $41.44\%$). Notice that despite the performance of CAP in some cases (such as weather) is slightly decreased, these corruptions will not appear in practical CT scans.

\paragraph{Visualization}
Figure \ref{img:interpretability} indicates that our CAP provides a sharper and more visually coherent gradient-based interpretability for perturbed CT scans. It is obvious that vanilla trained model cannot learn robust features from adversarial samples (second column). We alleviate the local omission of previous AT methods for large-scale infection (last row of the first two columns). On further investigation, the model trained by CAP highlights the contour of lung cavity and eliminates the attention outside the lung (last row of the last two columns). It proves our method may indeed learn the prior structure information and achieve a visual robustness. More figures are provided in Appendix D.4.

\paragraph{Resource Requirement}
To measure the resource requirement of different AT frameworks, we compute the seconds per training epoch for several high-accuracy baselines. In Visformer, the training time of PGD, TRADES, AWP, FAT and CAP are $\{$102.53, 105.49, 117.26, 135.09, 113.49$\}$. In Deit, the values of them are $\{$80.86, 119.75, 83.36, 111.39, 82.73$\}$. The results show that our algorithm only needs medium-sized computational resource when achieving the best performance.

\begin{table*}[th]
\centering
\caption{Clean/Adversarial performance of models trained with MosMed-L. We use both MosMed-L and MosMed-M for evaluation.}
\scalebox{0.84}{
\begin{tabular}{cc|ccc|ccc|ccc}
\toprule

\multicolumn{11}{c}{MosMed-L}\\
\midrule
        & & & AT & & & TRADES & & & CAP (Ours) & \\
        \midrule
        Class & Infection & Accuracy & Specificity & F1 Score & Accuracy & Specificity & F1 Score & Accuracy & Specificity & F1 Score \\
        \midrule
        0 & Healthy & 93.8/90.1 & 97.0/94.6 & 85.9/77.6 & \textbf{97.2}/89.0 & \textbf{98.4}/93.1 & \textbf{93.9}/75.8 & 96.2/\textbf{90.4} & 98.3/\textbf{95.0} & 91.5/\textbf{78.0} \\
        1 & $0<x<25$ & 89.2/82.7 & 79.1/70.2 & 91.6/86.6 & \textbf{94.9}/81.7 & \textbf{91.8}/73.5 & \textbf{95.9}/85.4 & 94.2\textbf{/85.7} & 88.7/\textbf{75.8} & 95.4/\textbf{88.8} \\ 
        2 & $25<x<50$ & 95.2/91.7 & 98.7/97 & 76.2/57.8 & \textbf{97.8}/92.2 & 99.1/96.4 & \textbf{90.2}/63.0 & 97.4/\textbf{94.0} & \textbf{99.1/97.6} & 87.9/\textbf{71.0} \\ 
        3 & $50<x<75$ & 98.0/97.3 & 99.6/99.2 & 71.1/61.5 & 98.7/96.6 & 99.5/98.4 & 83.7/55.8 & \textbf{98.8/97.5} & \textbf{99.8/99.1} & \textbf{84.0/65.9} \\
        4 & $75<x<100$ & 100/100 & 100/\textbf{100} & 100/100 & 100/100 & 100/100 & 100/\textbf{100} &\textbf{100/100} & \textbf{100/100} & \textbf{100}/80 \\
        Total & - & 88.1/80.9 & - & - & \textbf{94.3}/79.7 & - & - & 93.3/\textbf{85.6} &-&-\\
\midrule
\midrule
\multicolumn{11}{c}{MosMed-L $\to$ MosMed-M}\\

\midrule
        0 & Healthy & 50.0/54.4 & 52.8/66.8 & 27.1/11.2 & 26.7/15.5 & 6.0/2.2 & 37.6/24.6 & \textbf{76.5/74.0} & \textbf{98.7/95.9} & 3.0/NAN \\ 
        1 & $0<x<25$ & 40.5/36.5 & 94.6/90.6 & 12.5/5.1 & 41.3/30.0 & 96.7/75.1 & 12.4/2.0 & \textbf{61.4/58.8} & 1.4/0.5 & \textbf{75.9/74.0} \\ 
        2 & $25<x<50$ & 88.7/88.7 & 100/100 & NAN & 88.7/88.7 & 100/100 & NAN & \textbf{88.7/88.7} & \textbf{100/100} & NAN \\ 
        3 & $50<x<75$ &96.0/96.0 & 100/100 & NAN & 96.0/96.0 & 100/99.4 & NAN & \textbf{96.0/96.0} & \textbf{100/100} & NAN \\ 
        4 & $75<x<100$ &52.2/34.0 & 52.1/33.8 & \textbf{0.7/0.5} & 99.8/99.8 & 100/100 & NAN & \textbf{99.8/99.8} & \textbf{100/100 }& NAN \\ 
        Total &-& 13.7/4.8 & - & - & 26.2/14.5 & - & - & \textbf{61.2/58.7}&-&- \\ 
\bottomrule
\end{tabular}
}
\label{tbl:mosmed}
\end{table*}

\subsection{Severity Classification on MosMed}
We then consider a more challenging severity classification task and the results are shown in Table \ref{tbl:mosmed}. In source data distribution (Lung window scans), the accuracy, specificity and F1 score of CAP for adversarial samples are improved consistently. In particular, it outperforms the SOTA method, standard AT, by a large margin of 4.7$\%$ of adversarial accuracy in total five classes. In class 0, 1 and 2, our clean accuracy slightly decreases since the trade-off \cite{zhang2019theoretically} still exists in medical data. 

Furthermore, when deploying previous techniques to data with larger distribution shift (i.e., Mediastinal window), their adversarial robustness deteriorates a lot. For example, the clean and adversarial accuracy in TRADES are surprisingly 68.1$\%$/65.2$\%$ worse than they perform in MosMed-L, while CAP narrows the gap by only 32.1$\%$/26.9$\%$ respectively. Note that there is a numerical explosion in some evaluations, such as F1 score from class 2 to 4. The underlying reason is that the lower contrast in mediastinal window images affects the recognition of large-area diffuse patchy infections. Fortunately, we can still distinguish positive and negative patients with a better F1 score (75.9/74.0 vs the highest 12.5/5.1), even if it could underestimate the infection volumes.

\subsection{Ablation Study}
We perform an ablation study for each component of our CAP framework, including self-guided filter (SGF), attention parameter regularization (APR) and hybrid distance metric (HDM). As shown in Table \ref{tbl:ablation}, each component of CAP is indeed effective. Firstly, applying our edge-preserving augmentation in vanilla TRADES improves both adversarial accuracy and generalization ability by $2.01\%$ and $2.73\%$. After attention parameter regularization, all the three metrics are improved especially adversarial accuracy on SARS-COV-2, which indicates the importance of the way to embed the prior information. We further validate the necessity of HDM and results suggest the effectiveness of incorporating all the three components. More ablations about the parameter-sensitivity and the comparison with consistency regularization \cite{tack2022consistency} are carried out in Appendix D.1 and D.2, respectively. 

\begin{table}[hbt]
\centering
\caption{Ablation study on different components of our method on SARS-COV-2 and COV-M with Visformer-T.}
\scalebox{1.0}{
\begin{tabular}{ccc|ccc}
\toprule
        SGF & AWR & HDM & Clean & PGD-10 & COV-M\\
        \midrule
         & & &88.73&77.67&47.38\\
        \checkmark& & &86.52&79.68&50.11\\
        \checkmark&\checkmark& &90.14&83.50&51.71\\
        \checkmark&\checkmark&\checkmark&92.96&85.92&53.08\\
\bottomrule
\end{tabular}
}
\label{tbl:ablation}
\end{table}

\section{Conclusion}
In this paper, we propose a contour attention preserving (CAP) framework to improve the model-wise adversarial robustness in COVID-19 CT classification. An implicit parameter regularization is first introduced to inject the lung cavity prior feature into vision transformer. We further optimize the min-max problem via a hybrid distance metric. Then we propose a new COVID-19 CT dataset to evaluate the robustness under distribution shift. Extensive experiments demonstrate our method significantly improves the adversarial robustness and interpretability in medical imaging diagnosis.

\section{Acknowledgements}
This work is supported in part by the National Nature Science and Foundation of China Grand No. 71801031, and in part by the Guangdong Basic and Applied Basic Research Foundation project, China, No. 2019A1515011962.

\clearpage

\bibliography{aaai23.bib}

\end{document}